\documentclass[12pt,preprint]{emulateapj}
\usepackage{amsmath, natbib, placeins,color, enumitem}
\usepackage{color}

\pdfoutput=1

\def\beq{\begin{equation}}
\def\eeq{\end{equation}}
\def\bey{\begin{eqnarray}}
\def\eey{\end{eqnarray}}
\newcommand{\gtorder}{\mathrel{\raise.3ex\hbox{$>$}\mkern-14mu
            \lower0.6ex\hbox{$\sim$}}}
\newcommand{\ltorder}{\mathrel{\raise.3ex\hbox{$<$}\mkern-14mu
            \lower0.6ex\hbox{$\sim$}}}

\begin{document}

\title{A New Method for Finding Point Sources in High-energy Neutrino Data}
\author{Ke Fang \& M. Coleman Miller}
\affiliation{Department of Astronomy, University of Maryland, College Park, MD, 20742}
\affiliation{Joint Space-Science Institute, College Park, MD, 20742}

\begin{abstract}

The {\it IceCube} collaboration has reported the  first detection of high-energy astrophysical neutrinos including  $\sim 50$ high-energy starting events, but no individual sources have been identified.  It is therefore important to develop the most sensitive and efficient possible algorithms to identify point sources of these neutrinos.  The most popular current method works by exploring a dense grid of possible directions to individual sources, and identifying the single direction with the maximum probability of having produced multiple detected neutrinos.  This method has numerous strengths, but it is computationally intensive and, because it focuses on the single best location for a point source, additional point sources are not included in the evidence.  
We propose a new maximum likelihood method that uses the angular separations between all pairs of neutrinos in the data. Unlike existing autocorrelation methods for this type of analysis, which also use angular separations between neutrino pairs, our method incorporates information about the point spread function and can  identify individual point sources.  We find that if the angular resolution is a few degrees or better, then this approach reduces both false positive and false negative errors compared to the current method, and is also more computationally efficient up to, potentially, hundreds of thousands of detected neutrinos.   

\end{abstract}

\maketitle

\section{Introduction}
\label{sec:introduction}

The era of high-energy neutrino astronomy has been inaugurated by the first detections of high-energy neutrinos by the IceCube collaboration \citep{Aartsen:2013jdh, Aartsen:2013bka, Aartsen:2014gkd, 2015PhRvD..91b2001A}, and rich astrophysical returns are promised by the existing, projected and future high-energy and ultrahigh-energy (UHE) neutrino experiments, such as the Antarctic Impulsive Transient Antenna \citep{2010arXiv1003.2961T}, ANTARES \citep{2011NIMPA.656...11A},  the Askaryan Radio Array \citep{2012APh....35..457A}, the Antarctic Ross Ice-Shelf ANtenna Neutrino Array (ARIANNA \citealt{Barwick:2006tg}),  the Cubic Kilometre Neutrino Telescope (KM3NeT\footnote{http://www.km3net.org}),   the ExaVolt Antenna (EVA \citealt{2011APh....35..242G}),the Giant Radio Array for Neutrino Detection (\citealt{Martineau-Huynh:2015hae}), IceCube-Gen2 \citep{2014arXiv1412.5106I}, and the JEM-EUSO Mission \citep{2013arXiv1307.7071T}. One key objective of all of these experiments is to search for the origins of the neutrinos. Searches for point-like sources using individual events in the 4-year IceCube data release find no departure from an isotropic background \citep{Aartsen:2014cva}.  Thus the identification of individual sources requires more data \citep{Ahlers:2014ioa}, and it could also benefit from more advanced search techniques.

The point source search method currently employed by the IceCube Collaboration  \citep{2011ApJ...732...18A, 0004-637X-779-2-132, 2015APh....66...39A, Aartsen:2014cva} is based on an unbinned maximum likelihood ratio test which uses the spatial, energy, and temporal information of individual events to construct likelihoods \citep{Braun, Braun1}.  To assess the degree of spatial clustering when the source direction is not known, candidate source positions are evaluated over a dense grid of directions on the sky.  For each position, a test statistic (TS) is constructed which compares the likelihood that some fraction of the observed neutrinos come from that position with the likelihood that all neutrinos are drawn independently from an isotropic distribution, incorporating for both likelihoods the detector sensitivity and angular resolution as a function of direction and energy.  The TS value is then compared with a distribution of TS values from scrambled data, which thus do not have point sources, to determine the significance of any detection.  Because this method considers only a single source in each separate test, it does not optimally include information from multiple sources, if they are present.  We call this approach the {\it single-source} method.  Note that the angular steps between candidate point source directions must be much smaller than the angular resolution, in order to avoid missing possible sources.  In practice this means that this method is quite computationally intensive.

Alternatively, a two-point autocorrelation method has been applied  to search for small-scale anisotropies in neutrino data (e.g. in \citealt{2012NIMPA.662S.216L, Adrian-Martinez:2014hmp, 2015APh....66...39A}). 
 In this  approach, a cumulative distribution is constructed using pre-selected bins of angular separation between event pairs.  The cumulative number of pairs below the angular separation in each bin is then compared with the distribution obtained from many independent scramblings of the data, which should therefore represent source-free populations.  The intent of this approach is to detect any deviations from isotropy, but it does not include specific information about the point spread function and cannot localize individual sources.

In this work, we propose a new statistical method that utilizes event pairs  to search for and localize point sources (we refer to this as  the {\it pair method}).  The pair method focuses on the distribution of the angular separations between the observed neutrinos, and is therefore designed to pick out multiple events from directions that are the same within the tolerance of the point spread function, which is the key signature of point sources.  
In contrast to the approach of the autocorrelation method, we use a differential and unbinned distribution that is optimized for the detection of point sources.  We find that for angular resolutions of a few degrees or better the pair method is both more accurate and considerably faster than the single-source method.

 In addition, when the evidence for point sources is strong, while the single-source method points only to the location of the  brightest source and the autocorrelation method can only determine the angular size that has the greatest anisotropy, the pair method can localize all the point sources with a high success rate. 

\section{Description of the methods} \label{sec:methods}

We have two tasks in our point-source analysis of neutrino data:
\begin{enumerate}[label*=\arabic*.]
\item Determine the strength of the evidence in favor of point sources. Thus the following two models must be compared, where we assume that we do not have any prior information about the directions to sources:
\begin{enumerate}[label=(\alph*)]
\item All of the detected neutrinos come from a diffuse, isotropic population.  Thus no individual source has produced more than one detected neutrino.
\item Some fraction of the neutrinos come from specific point sources, with the rest being from a diffuse isotropic population.  Thus at least one individual source has produced more than one detected neutrino.

\end{enumerate}
\item Localize the sources from which multiple neutrinos have been seen. 
\end{enumerate}

By the term ``isotropic population" we mean a distribution of event origins that is statistically indistinguishable from being isotropic.  It could be background noise, such as atmospheric neutrinos in case of TeV neutrino detections. It could also be a cosmological population of neutrinos from an isotropic distribution of sources; for example, if we see 100 neutrinos but they come from $10^6$ sources of comparable brightness, it is likely that no particular source produced more than one neutrino in the sample.  We note that both methods could easily be extended to consider a diffuse population that is {\it not} isotropic (for example, ultrahigh energy cosmic rays are expected to follow the distribution of large-scale structure at distances $\ltorder 100$~Mpc).  However, for neutrinos the assumption of isotropy is likely to be good.

In Section~\ref{sec:single-source} we review  the single-source method.  In Section~\ref{sec:autocorrelation} we discuss the autocorrelation method.   We find that there are circumstances
   in which it is almost as efficient  as the pair method in identifying evidence of point sources, but it has no capacity for localizing the sources.  Finally we introduce the pair method in Section~\ref{sec:pair}.

We assume that the probability that a neutrino comes from a direction  with zenith angle $\theta$ and azimuthal angle $\phi$ is $P(\theta,\phi)$, which is determined by the detector sensitivity and exposure distribution including possible obscuration by the Earth.  We also assume that the probability of measuring a direction $(\theta^\prime,\phi^\prime)$ to that neutrino is $Q(\theta^\prime,\phi^\prime|\theta,\phi)$. Note that $Q(\theta^\prime,\phi^\prime|\theta,\phi)$ depends on the angular resolution  $\sigma(\theta,\phi)$ of the experiment given the true direction of the neutrino; in our simulations we assume that $\sigma(\theta,\phi)$ has the form of a circular Gaussian, but this can be generalized easily.  For the purposes of illustration, in this example we do not include information about the energy or time of arrival of each neutrino, but as we describe later we expect that this information can be incorporated straightforwardly.

\subsection{Single-Source Method}\label{sec:single-source}
In the single-source method, for a candidate source location $\vec{x}_s$ the log likelihood is \citep{Braun}:
\beq\label{eqn:lnL_SS}
\ln {\cal L}(f, \vec{x}_s) = \sum_i \ln \left[f \, {\cal S}_i + (1-f) \,{\cal B}_i   \right]\; .
\eeq
Here $f$ is the unknown fraction of events that came from the source at $\vec{x}_s$. ${\cal S}_i$ is the signal probability density function (PDF) that event $i$ is seen from direction $\vec{x}_i$ given a true source direction $\vec{x}_s$; following \cite{Braun} we model ${\cal S}_i$ as a 2-dimensional Gaussian, ${\cal S}_i = \exp\left[{-(\vec{x}_i - \vec{x}_s)^2 / 2\sigma_i^2}\right]/(2\pi\sigma_i^2)$. ${\cal B}_i$ is the background PDF, which is the normalized detection probability at $\vec{x}_i$, ${\cal B}_i = P(\vec{x}_i)/\int P(\vec{x})d\vec{x}$.  
To compare the two models described at the beginning of this section, a test statistic ${\rm TS}_{\rm SS}$ is defined in the single-source method as:
\begin{equation}\label{eqn:TS}
{\rm TS}_{\rm SS}(\vec{x}_s) = 2 \ln \left[ \frac{{\cal L} (\hat{f},\vec{x}_s)}{{\cal L}(f=0)} \right]
\end{equation}
where $\hat{f}$ is the value of $f$ that maximizes the total likelihood, and ${\cal L} (f=0)$ corresponds to an isotropic background.  The final ${\rm TS}_{\rm SS}$ is determined by maximizing ${\rm TS}_{\rm SS}(\vec{x}_s)$ over all directions $\vec{x}_s$.  Because the statistical distribution of $TS_{\rm SS}$, when there is no source present, is not in general an easily-computed function, the no-source ${\rm TS}_{\rm SS}$ distribution is computed in practice by determining ${\rm TS}_{\rm SS}$ for multiple independent scrambled versions of the data set \citep{Aartsen:2015yva}.  In our simulations we simply generate many sets of synthetic data without any individual sources to determine the ${\rm TS}_{\rm SS}$ distribution.  Note that the spacing of the directions $\vec{x}_s$ to be tested must be considerably finer than the angular resolution, to avoid missing the highest-likelihood location; for example, if the angular resolution is $1^\circ$ then a spacing of $0.1^\circ$ or smaller appears to be necessary.   Thus at this angular resolution several million possible source directions must be tried over the whole sky.

\subsection{Autocorrelation Method}\label{sec:autocorrelation}
The two-point autocorrelation method is most commonly used to detect intrinsic clusters within events. For data containing $N$ detected neutrinos, there are $N(N-1)/2$ unique pairs. For a range of angular separations with step size  $\Delta\Omega$, the autocorrelation function is defined as the cumulative number of pairs $(i,j)$ that have an angular separation $\alpha_{ij}$ smaller than  a given angle $\Omega$, 
\beq
N(\Omega) = \sum_{i,j>i} \,\mathcal{H}(\Omega-\alpha_{ij}).
\eeq
where $\mathcal{H}$ is the Heaviside step function. The optimal size of the angular steps needs to be pre-determined by pseudo-experiments that generate a randomized sky with the same detector configuration, because decreasing the step size enhances the angular resolution of the method but reduces the sensitivity due to the larger number of trials.  Reference data should  be generated by scrambling the data themselves or by Monto Carlo simulations.  In each  $\Omega$ bin, the reference autocorrelation function  $N(\Omega)_{\rm ref}$ provides the pair number expected if events are from an isotropic background.  Applying Poisson statistics, the probability of an excess or a deficit $p(\Omega)$ is determined  by $N(\Omega)$ and $N(\Omega)_{\rm ref}$ in that bin. The test statistic is then defined as the maximum of $p(\Omega)$ from all available bins:
\beq
{\rm TS}_{\rm AC} = \max(p(\Omega))
\eeq

Finally, to correct the trial factor due to the binning of the angular scales within which an anisotropy signal is sought, a large number of realizations of a randomized sky needs to be generated from an isotropic background and analyzed in the same way that we would analyze data. The fraction of realizations that produces a ${\rm TS}_{\rm AC}$ at least as large as seen in the data determines the significance of the signal. 

\subsection{Pair Method}\label{sec:pair}
The pair method also focuses on the angular separations between pairs of neutrinos.  Each pair either shares the same origin, or comes from different sources. Thus the $N(N-1)/2$ distinct pairs can be divided into a fraction $f_{\rm pair}$ that are same-source pairs, and a fraction $1-f_{\rm pair}$ that are not.

For the same assumptions about $P(\theta,\phi)$ and $Q(\theta^\prime,\phi^\prime|\theta,\phi)$ that we used for the single-source method, we can construct an angular separation probability distribution for an individual point source, $A_{\rm point}$, and for isotropic diffuse sources, $A_{\rm diff}$.  To determine $A_{\rm point}$ we repeatedly select the direction to a source using $P(\theta,\phi)$, then draw many neutrinos from that source based on $Q(\theta^\prime,\phi^\prime|\theta,\phi)$ and thus assemble a histogram of point-source angular separations $\alpha_{ij}$ for neutrinos $i$ and $j$ that is properly weighted by the detector sensitivity $P(\theta,\phi)$.  Similarly, we produce $A_{\rm diff}$ by selecting a large number of neutrino directions using $P(\theta,\phi)$, and determining the observed direction using $Q(\theta^\prime,\phi^\prime|\theta,\phi)$, which therefore produces a histogram of diffuse angular separations.  
Considering that neutrinos could have different angular resolutions due to their arrival directions or event types, 
we rescale the angular separations by a sum in quadrature of the angular resolutions $\sigma_i$ and $\sigma_j$ of the two events in a pair: $\bar{\alpha}_{ij} \propto \alpha_{ij} /\sqrt{ \sigma_i^2+ \sigma_j^2}$.

The log likelihood of the data given the model can then be written as  
\bey\label{eqn:lnL}
\ln{\cal L}(f_{\rm pair})&=&\sum_{i,j>i}\,\ln  [f_{\rm pair}\,A_{\rm point}(\bar{\alpha}_{ij}) \\  \nonumber
&+& \,(1-f_{\rm pair})\,A_{\rm diff}(\bar{\alpha}_{ij})]\; .
\eey

We then define a test statistic ${\rm TS}_{\rm pair}$ in the same way as in Equation~(\ref{eqn:TS}), but without a dependence on source location.  As with the single-source method, the distribution of ${\rm TS}_{\rm pair}$ when there are no point sources must in general be computed numerically.

\section{Results}
\subsection{Numerical Setup}
The detection probability and point-spread function as a function of direction will vary from experiment to experiment, and will also change with time for a given experiment. Here we compare the pair method with the single-source method using illustrative assumptions for $P(\theta,\phi)$ and $Q(\theta^\prime,\phi^\prime|\theta,\phi)$, but we note that the results can easily be extended to more general cases with realistic sensitivity and angular resolution. 

We assume that the probability of detection of a neutrino is $P(\theta, \phi) = \cos\theta$, so that the detection probability goes to zero for neutrinos at the horizon.  As before, we adopt a point spread function that is a 2-dimensional Gaussian, $Q(\vec{x}^\prime|\vec{x}) = \exp\left(-(\vec{x}-\vec{x}^\prime)^2/2\sigma_{\vec{x}}^2\right)/(2\pi\sigma_{\vec{x}}^2)$. The angular resolution is set to be $\sigma (\theta,\phi)=0.01(2-\cos\theta)$ radians; that is, the angular resolution at the zenith is twice as good as it is at the horizon.  

We further assume that sources can be seen perfectly out to an abrupt edge at $R_{\rm max}=2$~Gpc (corresponding to redshift  $z\sim 0.5$).  For simplicity we ignore the evolution of source emissivity over redshift.  We assume that the point sources have a uniform probability of being anywhere in space, with a number density $n_s$.  Thus if $n_s$ is large, the neutrinos from most of those sources will be indistinguishable from neutrinos from a diffuse isotropic population. The source number density $n_s$ is left as a free parameter, and we assume that the source number in the sky follows a Poisson distribution with a mean value $n_s\,4\pi R_{\rm max}^3/3$.

Given our choice of angular resolution (which is $\sim0.6^\circ$ at best), for the single-source method we scan the sky with $0.1^\circ\times 0.1^\circ$ bins in $(\theta,\Phi)$; for the autocorrelation method we use angular  steps with an increment  of  $0.1^\circ$ from $0.1^\circ$ to $3^\circ$. 

For the single-source method and the pair method, the best-fit likelihood is obtained by numerically maximizing $\ln{\cal L}$ (in equation~\ref{eqn:lnL_SS} and \ref{eqn:lnL}) over  $f$ in the range $f\in[0, 1]$. 

\subsection{Identification of Evidence}

\begin{figure}
\includegraphics[scale=0.45]{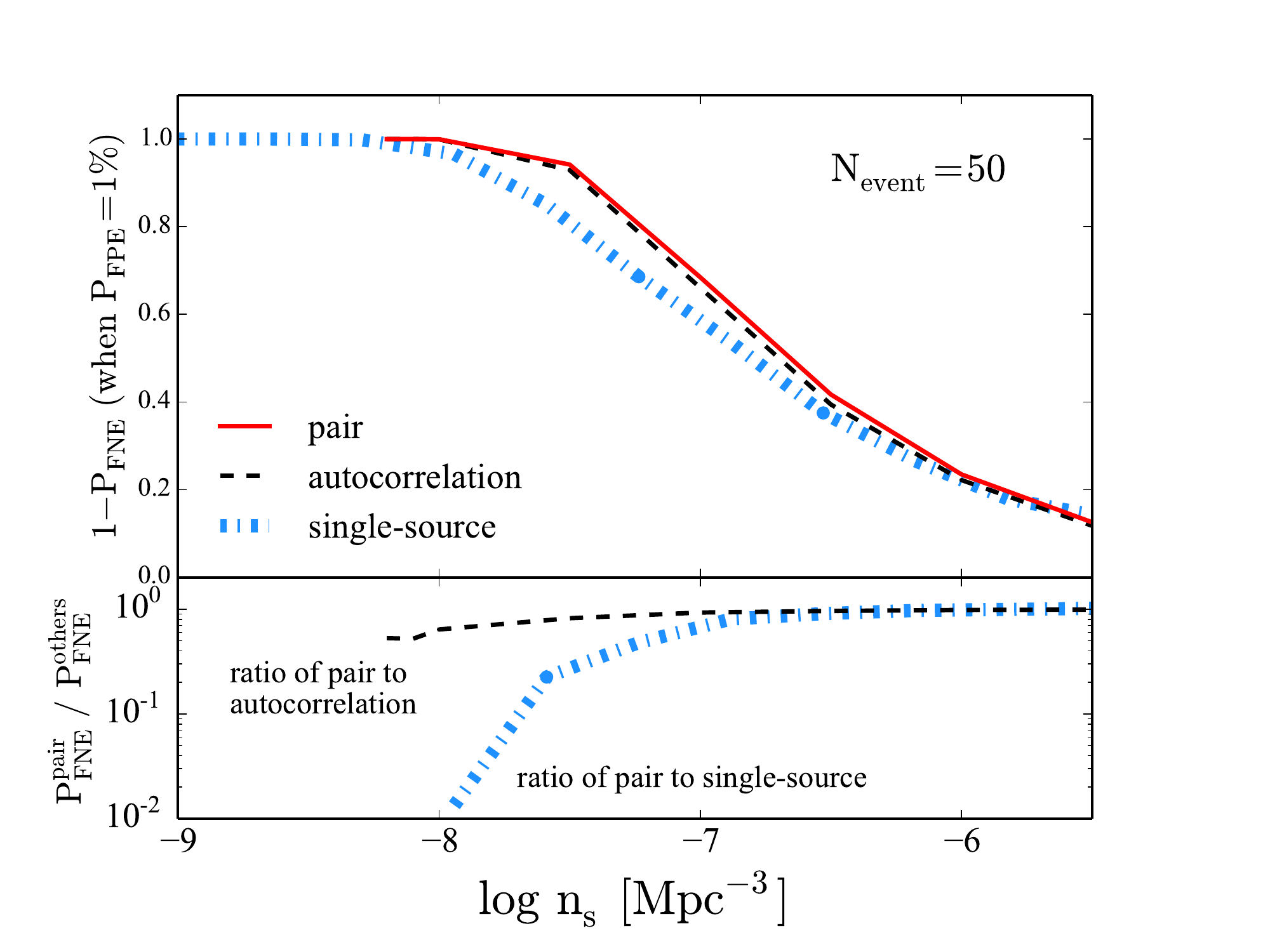}
\caption{\label{fig:ratio} Top: 
The fraction of realizations successfully detected,  $(1- P_{\rm FNE})$ (with $P_{\rm FNE}$ being the false negative error rates), of the pair method proposed in this work (solid red line), the single-source method developed in \cite{Braun} (dash-dotted cyan line), and  the autocorrelation method used in e.g. \citet{Adrian-Martinez:2014hmp} and \citet{2015APh....66...39A} (dashed black line), when the false positive error rates of all methods reach  $1\%$, for a range of source number densities and a fixed neutrino number $N_{\rm event} = 50$. Bottom: The ratios of the false negative error rate of the pair method to that of    the single-source method (dash-dotted cyan line) and the autocorrelation method (dashed black line). See text for details.  In all tests for which the number density of sources is less than $n_s=10^{-6.5}\,\rm Mpc^{-3}$ and $N_{\rm event}=50$, the pair method produces a significantly smaller false negative rate than the single-source method.  The detection limit is reached for $n_s\sim 10^{-6.5}\,\rm Mpc^{-3}$, so for higher number densities a successful differentiation of sources from background would require more events or (possibly) better angular resolution. 
This figure demonstrates that when there are enough events to detect point sources, the pair method does better than the autocorrelation method and significantly better than the single-source method, and that with stronger evidence in favor of point sources, the advantage of the pair method is increased.
}
\end{figure}

A good way to measure the quality of a search method is to determine the rate of false positive errors (FPE) and false negative errors (FNE). A FPE occurs when the neutrinos actually come from a diffuse background, but the TS value exceeds the chosen threshold for detection and thus there is a false report of a signal.  Similarly, a FNE happens when the data were produced by a point source population, but the TS value is below the chosen threshold and thus the isotropic background is incorrectly favored. In some circumstances, it might be desirable to minimize false positive errors (for example, when one wants to be confident about a claimed detection), while in others it might instead be preferable to minimize false negative errors (for example, when multi-messenger followups are planned and it is undesirable to miss a real source).

Figure~\ref{fig:ratio} compares the probabilities of FPE and FNE for the pair method with those for the single-source method and the autocorrelation method in tests with sources in  a range of number densities.  We fix the event number to be $N_{\rm event} = 50$ to correspond approximately to the number of detected high-energy starting events   in the 4-year IceCube data.  
In the top panel, the solid red, dashed black, and dash-dotted cyan lines indicate the FNE rates of the pair method, the autocorrelation method, and the single-source method, respectively, for a TS threshold that results in an FPE rate of $1\%$ in null tests. In the bottom panel, the dashed black line shows the ratio between the pair method and the autocorrelation method, while the dash-dotted cyan line   corresponds to the ratio between the pair method and the single-source method.  
The ratios of FNE rates of the two methods when their FPE rates reach $10\%$ and $0.1\%$ follow a similar trend to the 1\% results.  This figure shows that with 50 events and degree-level angular resolution, the pair method has a significantly lower FNE rate than the single-source method for $n_s<10^{-6.5} \,\rm Mpc^{-3}$.   Compared to the autocorrelation method, the pair method has up to $50\%$ smaller relative error probability  for the number densities that we tested (that is, $P_{\rm FNE}^{\rm Pair} / P_{\rm FNE}^{\rm AC} \geq 50\%$). 
  For larger $n_s$ and $N_{\rm event} = 50$, none of the methods can differentiate sources from an isotropic background. The number of events required for a given FPE or FNE rate scales with the source number density roughly as $N_{\rm event}\propto n_s^{0.55}$ for the pair method and the single-source method, but increases faster for the autocorrelation method (also see the subplot of Fig~\ref{fig:Nneed}). 
    Overall, we find that as the strength of evidence in favor of point sources increases, so does the efficiency of the pair method relative to the single-source method and the autocorrelation method.  We find, however, that the advantage of the pair method appears to diminish as the angular resolution becomes worse, and that the three methods appear comparable for angular resolutions around 10 degrees.

\begin{figure}
\includegraphics[scale=0.45]{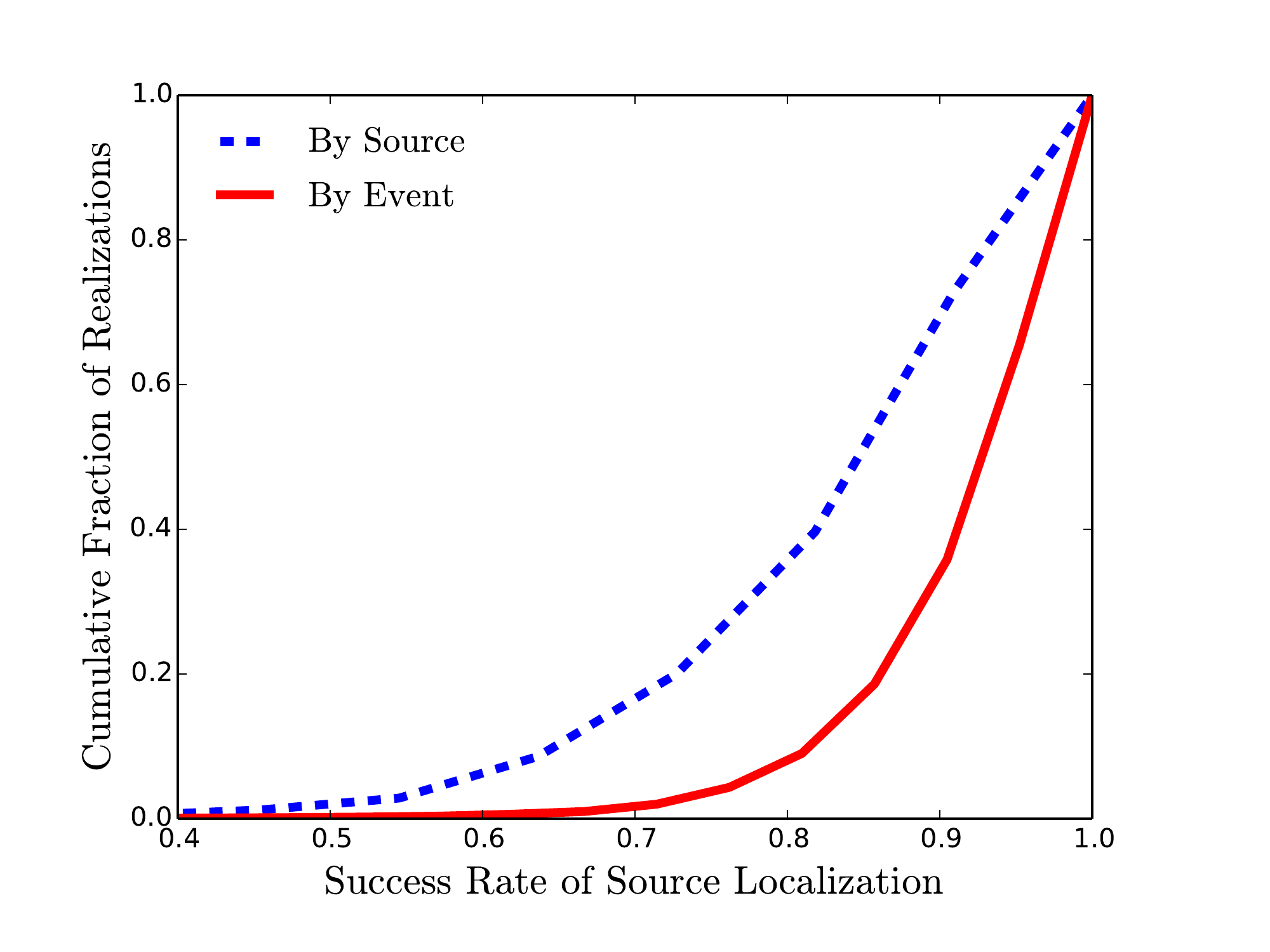}
\caption{\label{fig:localization} 
Cumulative distribution of the success rate of source localization by the pair method.  We performed $5\times10^5$ independent tests. In each test, we generated 50 events from point sources with a number density of $n_s = 10^{-8}\,\rm Mpc^{-3}$. We list event pairs in decreasing order of $E(\bar{\alpha}) = A_{\rm point}({\bar\alpha})/A_{\rm diff}({\bar\alpha})$ until we see a sudden drop in $E(\bar{\alpha})$.  A success in source localization is defined as finding all the events, and only the events, from a source ({\it By Source}, dotted blue line) or finding at least one event that came from a source ({\it By Event}, solid red line) (see to Section~\ref{sec:localization} for more details). In both cases, 
the pair method can localize all the sources from a given simulation with a high success rate.
}
\end{figure}

\subsection{Source Localization}\label{sec:localization}

As stated in Section~\ref{sec:methods}, one important function of the search method is to localize sources.   The pair method can achieve this by listing the best candidate positions for individual sources.  Essentially, for a given pair of events $i$ and $j$, the evidence in favor of these arising from the same point source compared to the events not arising from the same source is $E(\bar{\alpha}_{ij}) = A_{\rm point}({\bar\alpha}_{ij})/A_{\rm diff}({\bar\alpha}_{ij})$. The event clusters that contain the pairs with the largest values of $E(\bar{\alpha})$ map out the sources. The  coordinate of a source can then be derived from the event cluster, for example by applying the single-source method to the subregion.

Figure~\ref{fig:localization} presents the success rate of source localization by the pair method. We performed $5\times10^5$ independent tests to compute the cumulative distribution of the success rate. In each test, we generate 50 events from point sources with $n_s = 10^{-8}\,\rm Mpc^{-3}$; we choose this number density to guarantee strong evidence in favor of point sources.  We then sort $E(\bar{\alpha})$ for all event pairs in decreasing order. The top $n$ pairs are selected until the $(n+1)$th pair displays a sudden drop compared with the average of the previous $n$ entries: $E(\alpha_{n+1}) /E(\alpha_{n}) < \eta\,\left(\sum_{i < n}\,  E(\alpha_{i+1}) /E(\alpha_{i})\right)/n $. $\eta$ is determined empirically, and setting $\eta=0.1 -0.5$ led to similar results in the current test. We define two types of  success rates of source localization:  1) {\it By source}: a localization is successful if and only if all the events of a source are found by the method. The success rate is obtained by dividing the number of successfully identified sources by  the larger   of  the actual source number and the source number found by analysis. 2) {\it By event}: a localization is successful if at least one of the events of a source is found by the method. The success rate is computed by dividing the number of successfully identified events by the larger of the actual same-source events and the same-source events found by analysis.  Figure~\ref{fig:localization} shows that for both definitions, the pair method can localize all the sources with a high success rate. 

\subsection{Event Number Needed for Detection}
\begin{figure}
\includegraphics[scale=0.45]{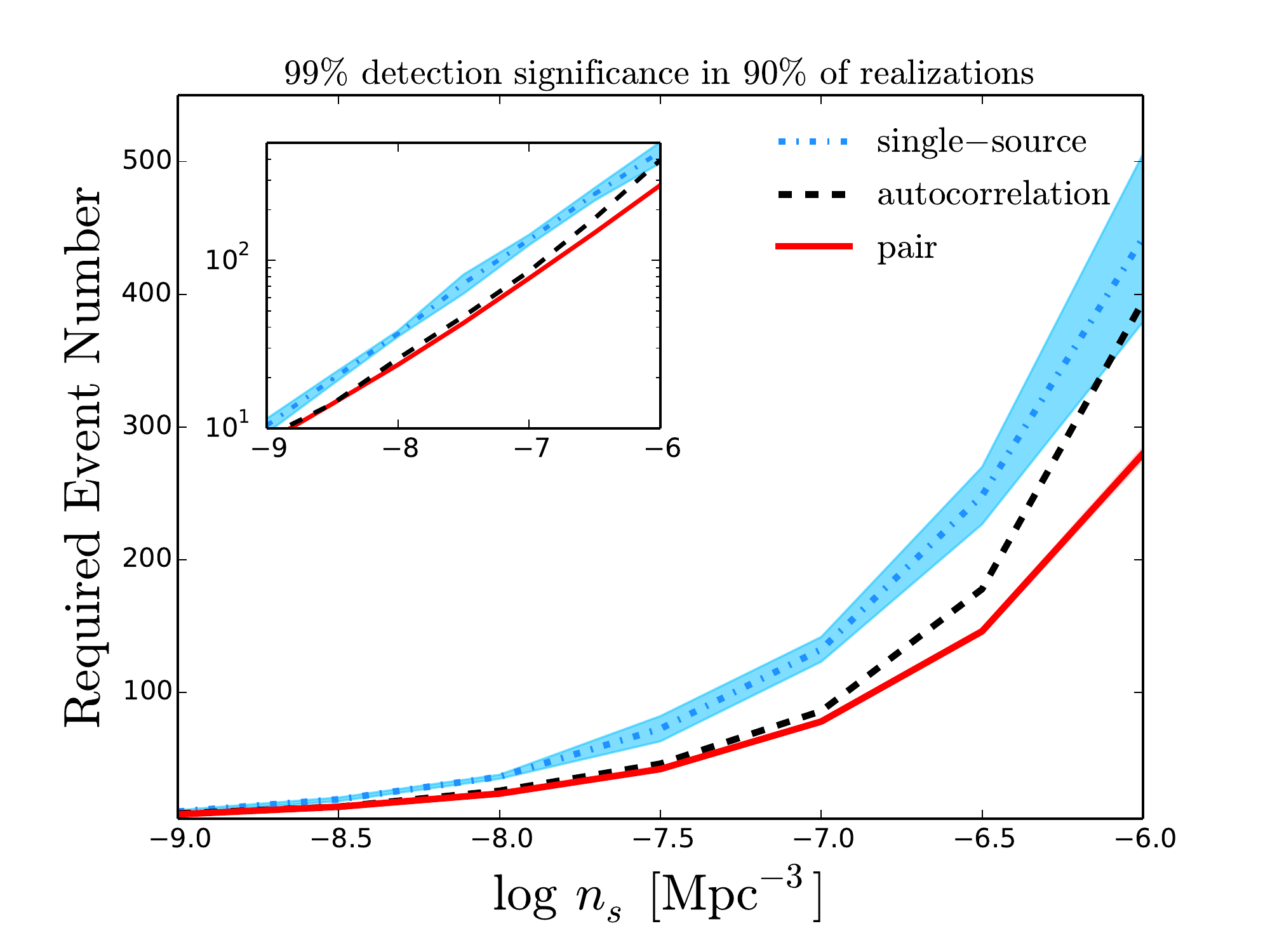}
\caption{\label{fig:Nneed} Number of events needed by the pair method, the autocorrelation method, and the single-source method to discover point sources, as a function of the number density of sources.  Here our criterion is that in 90\% of realizations, the null hypothesis (no point sources) can be rejected at the 99\% confidence level.  The detection probability and angular resolution of the simulated detector are the same as they were for the analysis shown in Figure~\ref{fig:ratio}.     
The subplot shows that at a given false positive rate,  the number of events required to reach a desired false negative rate scales roughly as $N_{\rm event}\propto n_s^{0.55}$ for the pair method and the single-source method, but evolves faster for the autocorrelation method. 
The larger error bars associated with the single-source method come from a less accurate calculation  due to computational resource limits, in which we allowed a rejection of null hypothesis at the $99\%$ confidence level but in $90\pm2\%$ of realizations. 
The pair method can significantly reduce the required event number compared to both the autocorrelation method and the single-source method, particularly for larger source number densities. }
\end{figure}

In Figure~\ref{fig:Nneed} we show the number of events needed to establish that there are point sources, as a function of the number density of neutrino sources.  Our criterion is that in 90\% of realizations with individual sources at a given number density, the TS must be at the 99th percentile or higher of the TS distribution produced by null tests (corresponding to a p-value of $0.01$).  The pair method requires significantly fewer neutrinos than the single-source method. 
The pair method and the autocorrelation method require similar number of events for $n_s \leq 10^{-8}\,\rm Mpc^{-3}$. As $n_s$ grows, the number of events needed by the autocorrelation method increases faster than that needed by the pair method. The advantage of the pair method becomes more evident for $n_s\geq 10^{-6.5}\,\rm Mpc^{-3}$.
  Similarly, we find that the number of neutrino detections needed to reject a given source number density, when the simulated neutrino directions are drawn from an isotropic distribution, is significantly less for the pair method than for the other methods.

For context, we note that many candidate neutrino source types have been proposed with a wide range of possible source number densities (e.g.,  massive galaxy clusters have $n_s \sim 10^{-7}-10^{-5}\,\rm Mpc^{-3}$\citep{Berezinsky:1996wx},  radio-loud active galactic nuclei (AGN)  have $n_s\sim 10^{-5}-10^{-4}\,\rm Mpc^{-3}$\citep{Tueller:2007rk}, and starburst galaxies have $n_s\ge 10^{-4} \,\rm Mpc^{-3}$  \citep{Thompson:2006np}). The sources might also be transient (e.g., low-luminosity gamma-ray bursts \citep{Murase:2013ffa}, newborn pulsars \citep{Fang:2013vla}, or giant AGN flares \citep{Farrar:2008ex}), and these have rates per volume ranging from $10^{-9} - 10^{-5}\,\rm Mpc^{-3}~yr^{-1}$. 
For the highest energy neutrinos from the interaction between ultrahigh-energy cosmic rays (UHECR) and the cosmic microwave background, the lack of significant clustering in the arrival directions of UHECRs implies a lower bound on the density of neutrino sources of $n_s\ge (0.06-5)\times10^{-4}\,\rm Mpc^{-3}$  \citep{2013JCAP...05..009P}. For reference, \cite{Ahlers:2014ioa} found that with the help of source associations, the lack of identified point sources suggests $n_s\gtorder 10^{-6}\,\rm Mpc^{-3}$.
Figure~\ref{fig:Nneed} indicates that at $n_s=10^{-6}\,{\rm Mpc^{-3}}$, the pair method is already capable of reducing the required event number by $\sim 100$.  Considering that the full configuration of IceCube detects $\sim 10$ high-energy starting events per year and $\sim 10$ muon neutrino events per year \citep{Aartsen:2015yva}, the pair method could save years of observation time 
for the purposes of source identification and localization.

\section{Discussion} 
We have introduced a new method to search for unknown point sources in high-energy neutrino data, which focuses on the angular separation between pairs of neutrinos.  This method requires no prior information about the source location. For synthetic data observed with degree-level angular resolution we find that the pair method is more efficient than the currently-used single-source method \citep{Braun,Aartsen:2014cva}, both in determining that there are point sources when there are, and in setting lower limits on the source number density when the simulated neutrinos come from an isotropic background.  
In addition, the pair method is efficient at localizing sources. This is a unique feature compared with the autocorrelation method, which is most commonly used to detect intrinsic clusters in data. 

So far we have focused on the spatial dependence of likelihoods, and all tests in this work assume that events from sources and from background are indistinguishable and  have exactly same properties, including energy, arrival frequency, and source number density.
However, we expect that it will be straightforward to extend our method using the time and energy dependence of events. This can be done by redefining $\ln{\cal L}(f_{\rm pair})$ in Equation~(\ref{eqn:lnL}) to be 
\beq
\begin{array}{l}
\sum_{i,j>i}\ln[f_{\rm pair}\,A_{\rm point}(\bar{\alpha}_{ij})\,{\cal E}_{\rm point}(E_i, E_j)\,{\cal T}_{\rm point}(t_i, t_j)\\
+(1-f_{\rm pair})\,A_{\rm diff}(\bar{\alpha}_{ij})\,{\cal E}_{\rm diff}(E_i, E_j)\,{\cal T}_{\rm diff}(t_i, t_j)]\; ,
\end{array}
\eeq
where ${\cal E}_{\rm point}$ (${\cal E}_{\rm diff}$) and ${\cal T}_{\rm point}$ (${\cal T}_{\rm diff}$) describe the probability of two events being from a point source (diffuse background) as a function of their energies and arrival times. The time dependence is critical for the study of transient sources,
while the energy dependence is crucial for distinguishing the atmospheric backgrounds from actual astrophysical sources in TeV data.  We will present a detailed study of the incorporation of time and energy dependence in a future paper.  

We also anticipate that our method can be extended in other ways.  For example, if the source itself is not pointlike compared with the detector angular resolution (e.g., if the source is a nearby galaxy cluster), it should be possible to treat the effective angular resolution by adding the detector resolution in quadrature to the angular size of the source.  In the construction of $A_{\rm point}$ this could even be done as a function of redshift, for a given model of the linear size of the sources.  Similarly, for particles such as high-energy cosmic rays, whose paths are deflected somewhat by intergalactic or interstellar magnetic fields, the effective angular resolution could include the spread in directions as a function of energy.  Thus we expect that our approach could also be used for ultrahigh energy cosmic rays and gamma-rays as well as for neutrinos.

To compare the computing speeds of the single-source method and the pair method, we did a simple test by finding the test statistic of a given data set containing 50 events, assuming the same  detector setup used in Figure~\ref{fig:ratio}.  With a 2.9 GHz single-core processor, the single-source method took $930$~s whereas the pair method took 0.07~s.  The single-source method could potentially be optimized by ignoring events that are too far away from the assumed source location in a maximum likelihood calculation. 
 Even with an optimization factor $\sim 30$, which can be realized if one only considers the nearest 3 pixels, the pair method is still hundreds of times faster than the single-source method.  For larger numbers of neutrinos, the $N^2$ scaling of our method compared with the $N$ scaling of the single-source method might reduce the computational advantage, but even so our method should be as fast or faster up to $\sim (930/0.07)\times 50\approx 600,000$ events for the assumptions in our simulations. 
 
In summary, we believe that using the angular separations of pairs of neutrinos to detect point sources has significant advantages over current techniques.  This approach enhances the speed and accuracy of point source detection and localization and the setting of lower limits on source number density.  Finally, we anticipate that the pair method will also be useful in the analysis of ultrahigh-energy cosmic rays and gamma-rays.

\acknowledgments
We thank Markus Ahlers, Erik Blaufuss, Kara Hoffman, Ryan Maunu, Naoko Kurahashi Neilson, John Pretz, and Greg Sullivan for helpful comments.  We also acknowledge the University of Maryland supercomputing resources (http://www.it.umd.edu/hpcc) and  The Maryland Advanced Research Computing Center (https://www.marcc.jhu.edu) made available for conducting the research reported in this paper.  K.F. acknowledges the support of a Joint Space-Science Institute prize postdoctoral fellowship.  

\bibliography{FM16} 
\end{document}